\documentclass[12pt]{iopart}

\usepackage{arxiv}

\usepackage{amstext}
\usepackage{iopams}
\usepackage{graphicx}
\usepackage[colorlinks=true,citecolor=blue,linkcolor=blue]{hyperref} 

\newcommand{\dd}[0]{{\text{d}}}

\begin{document}

\title[Spontaneous waves in muscle fibres]{Spontaneous waves in muscle fibres}
\author[S.G.]{Stefan G\"unther and Karsten Kruse}
\address{Saarland University, Department of Theoretical Physics, 66041 Saarbr\"ucken, Germany}
\address{Max Planck Institute for the Physics of Complex Systems, N\"othnitzer Str.~38, 01187 Dresden, Germany}
\ead{stefan.guenther@physik.uni-saarland.de}
\ead{k.kruse@physik.uni-saarland.de}

\begin{abstract}
Mechanical oscillations are important for many cellular processes, e.g., the beating of cilia and flagella
or the sensation of sound by hair cells. These dynamic states originate from spontaneous oscillations of 
molecular motors. A particularly clear example of such oscillations has been observed
in muscle fibers under non-physiological conditions. In that case, motor oscillations lead to 
contraction waves along the fiber. By a macroscopic analysis of muscle fiber dynamics we find that
the spontaneous waves involve non-hydrodynamic modes. A simple microscopic model of sarcomere
dynamics highlights mechanical aspects of the motor dynamics and fits with the experimental observations. 
\end{abstract}

\maketitle

\section{Introduction}

Oscillations are a common phenomenon in biological systems. Even on a cellular level oscillatory 
dynamics is widespread~\cite{krus05}. Examples are provided by circadian rhythms through which
cells anticipate the changes of day and night~\cite{gold02}, by the Min-oscillations in the bacterium 
\textit{Escherichia coli} that help to select the cell center as the division site~\cite{roth06}, by hair bundle
oscillations in auditory hair cells~\cite{mart99}, and by the beating patterns of cilia and flagella that
are crucial for the transport of mucus and propel microorganisms~\cite{bray01}.

The latter are examples of mechanical oscillations that involve molecular motors~\cite{bray01,howa01,albe02}. 
These proteins are able to transform chemical energy into mechanical work. The chemical energy is 
provided by the hydrolysis of Adenosine-Tri-Phosphate (ATP) which results in Adenosine-Di-Phosphate 
(ADP) and inorganic phosphate (P$_i$). Motor proteins can catalyze this reaction and concomitantly 
undergo conformational changes. By this motion they can translate directionally along polar protein
filaments and transport cargos or generate forces. Examples are myosin motors that interact with actin 
filaments and kinesins or dyneins that interact with microtubules. The polarity of actin filaments and
microtubules results from structural differences between the two ends of a filament and a microtubule
and determines the direction of motion of a motor. 

\begin{figure}[bp]
\includegraphics[width=0.6\textwidth]{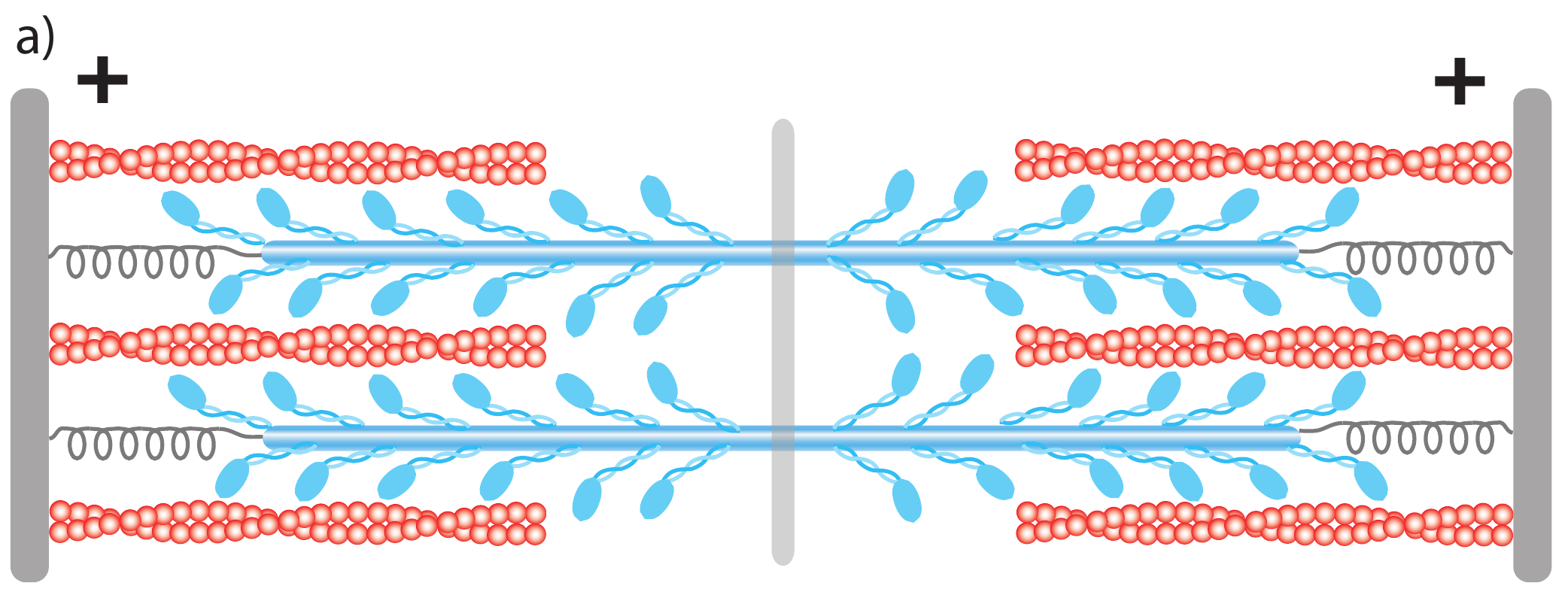} 
\caption{\label{fig:schemasarcomere}Illustrations of a muscle sarcomere. Attached with their plus ends
to the Z-discs, actin filaments are depicted in red. Interdigitating with the actin filaments are bipolar Myosin-II 
filaments. The motor heads can bind to the actin filament. Upon activation they move towards the actin
filaments' plus ends and contract the structure. The central structural element is the M-line. Together with the
titin molecules, which are represented as springs that go through the hollow myosin filaments, the Z-discs and
still other passive elements, they provide structural integrity to sarcomeres.}
\end{figure}On theoretical grounds it had been suggested that ensembles of molecular motors connected to 
elastic elements can generate spontaneous oscillations through a Hopf-bifurcation~\cite{juli97}. It has 
been proposed that such
mechanical oscillations are at the origin of flagellar and ciliary beats~\cite{brok75,cama99,cama00}
and of mitotic spindle oscillations during asymmetric cell divisions~\cite{gril05}. A striking evidence
of spontaneous oscillations caused by molecular motors is provided by oscillations of muscle fibers
that have been studied in the past 20 years by Ishiwata and 
collaborators~\cite{okam88,anaz92,ishi93a,ishi93b,yasu96,fuku96,fuji98,sasa05}. Muscle fibers are 
chains of sarcomeres, the elementary force generating units of skeletal and cardiac muscle~\cite{bray01},
see Fig.~\ref{fig:schemasarcomere} for an illustration. 
In a sarcomere actin filaments and filaments consisting of many myosin motor molecules are arranged
in such a way that activation of the motors leads to contraction of the structure. Its integrity is maintained
by passive elastic elements. 

Spontaneous sarcomere oscillations are observed under constant, but non-physiological chemical
conditions~\cite{okam88}. Remarkably, the oscillations exists even in reconstituted muscle
fibers that contain only essential structural elements like actin, myosin, and scaffolding proteins, but 
no regulatory elements~\cite{fuji98}. Depending on the external conditions and applied forces, the 
oscillations of sarcomeres in a muscle fiber can be either synchronous or 
asynchronous~\cite{fuku96,yasu96}. Under appropriate conditions, contraction waves traveling along 
the muscle fiber can be observed~\cite{okam88,anaz92,sasa05}. 

Theoretical work on this system has so far focused on synchronization effects of chains of coupled 
Hopf-oscillators~\cite{vilf03}. In that analysis, synchronization effects due a global coupling of 
oscillatory elements through a mass have been studied. In particular, a rich phase diagram of 
synchronous and asynchronous states was found. In another study, gradients in sarcomere properties
where suggested to cause synchronization between adjacent sarcomeres which ultimately yields
coherent contraction waves~\cite{smit94}. The existence of such a gradient is currently not supported  
by experiments. A comprehensive theoretical account of the 
experiments by Ishiwata et al.~is currently not available.

We present in this work a study of the dynamics of muscle fibers using two complementary approaches.
Phenomenological descriptions of active polar gels provide a general framework for studying the dynamics 
of motor filament systems~\cite{krus03a,krus04,krus05a,zumd05} and have been successfully applied, 
for example, to describe essential features of the lamellipodia of crawling cells~\cite{krus06}. We will 
start therefore in the next section by developing a hydrodynamic theory of muscle fibers. As will be shown, 
this description fails to yield oscillatory solutions. In the subsequent section we will consider a microscopic 
model of muscle sarcomeres in which spontaneous oscillations are a consequence of load dependent 
detachment rates of motors from filaments. The corresponding model of a chain of sarcomeres is then 
shown to wholly reproduce the phenomenology of waves along muscle fibers. In the discussion, we compare 
the results of our theoretical analysis to experimental observations.

\section{Hydrodynamic description of muscle fibres}
\label{sec:hydrodynamic}
Hydrodynamics is a systematic approach to assess dynamic phenomena of spatially extended systems in the
limit of large wave-lengths and long time-scales. Formally, a hydrodynamic mode of wave-length $q$ relaxes 
with a characteristic time $\tau\sim q^{-2}$. In this limit, it is appropriate to make the assumption of local 
thermodynamic equilibrium~\cite{degr84}: In thought, the system is subdivided into volumes that are small 
compared to the large scale structures of interest and that are, at the same time, large enough to allow for
a thermodynamic (equilibrium) description. Based in this assumption, a free energy can be defined for the 
full system which is out of equilibrium. The change with time of the free energy can be expressed as a sum
of products of generalized forces and fluxes. Then, the fluxes are expressed in terms of the forces, where only
terms of linear order are considered. The resulting description is purely phenomenological and only depends on 
the modes under consideration as well as the symmetries of the system.

We will apply this approach to muscle fibers, which will be considered as one-dimensional one-component 
complex fluids. First, we have to identify the hydrodynamic modes of the systems. In general,
they are associated with conservation laws or broken continuous symmetries. The muscle fiber is an essentially 
one-dimensional
system with the order of the actin and myosin filaments being fixed. Consequently the system does not offer any broken 
continuous symmetry. There are three conserved quantities: the fiber mass, momentum and energy. In experiments,
the fluid surrounding the fiber essentially provides a heat bath such that temperature is constant and energy 
conservation is not an issue. Note, that this no longer holds for muscle, which heat up when active. Conservation 
of mass and momentum implies
\begin{equation}
\label{eq:massconserv}
\partial_t \rho + \nabla \cdot \rho v = 0 \quad, 
\end{equation}
where $\rho$ is the mass density along the fiber and $v$ its local velocity, and
\begin{equation}
\label{eq:momconserv}
\partial_t \rho v - \nabla \cdot  \sigma =  f_{\text{ext}} \quad.
\end{equation}
In this expression the stress $\sigma$ gives the momentum flux density $-\sigma$ and $f_\text{ext}$ is an externally
applied force. For the dynamic phenomena we consider, inertia is irrelevant, such that Eq.~(\ref{eq:momconserv})
reduces to a force-balance relation between internal stresses and external forces.

Under the assumption of local thermodynamic equilibrium, changes in the system's free energy can be expressed as
\begin{equation}
\frac{\dd}{\dd t}F=-\int \left({\sigma\:\partial_x v+r\:\Delta\mu}\right) \dd x \quad.
\end{equation}
It can thus be written as a sum of products, where in each product, a generalized flux is 
multiplied with its conjugate generalized force. In the present case, the fluxes are taken to be the stress $\sigma$
and the rate of ATP-hydrolysis $r$. The force conjugate to $\sigma$ is the rate of strain $\partial_x v$, the force
conjugate to $r$ the difference $\Delta\mu$ in chemical potentials of ATP and its hydrolysis products ADP and P$_i$,
$\Delta\mu=\mu_{\text{ATP}}-\left(\mu_{\text{ADP}}+\mu_{\text{P}}\right)$. Note, that since the density changes by relative
sliding of actin filaments and motor filaments, we assume that the system is infinitely compressible. Consequently, the 
pressure vanishes.

In the next step, the generalized fluxes are expanded in terms of the generalized forces up to linear order. To this end 
the fluxes have to be separated into their reactive and their dissipative components. The dissipative component of 
a flux has the same sign with respect to time-reversal as its conjugate flux, the reactive component has the opposite sign.
The corresponding changes in the free energy are thus, respectively, irreversible and reversible with respect to 
time-reversal. The respective components of a flux can be expanded only in terms of forces with the same behavior 
under time-reversal. With $\sigma=\sigma^r+\sigma^d$ and $r=r^r+r^d$, where the superscripts $r$ and $d$ 
discriminate between the reactive and the dissipative parts, respectively, we get
 \begin{eqnarray}
 \label{eq:redis1}
 r^\text{r} &=& -\zeta \: \partial_x v \\
 r^\text{d} &=& \Lambda \: \Delta\mu \\
 \label{eq:redis3}
 \sigma^\text{r} &=& \zeta \: \Delta\mu \\
 \label{eq:redis4}
 \sigma^\text{d} &=& \xi \: \partial_x v \quad.
 \end{eqnarray}
The phenomenological coefficient $\xi$ accounts for the effects of an internal friction in the muscle fiber, while $\Lambda$
determines the rate of ATP-hydrolysis given a difference in chemical potentials $\Delta\mu$. The coefficient 
$\zeta$ is a measure of the contribution
to the stress by active processes, i.e., the action of motors. The coefficients in front of the cross-terms must be equal
up to a sign due to the Onsager relations. 

Equations (\ref{eq:redis1})-(\ref{eq:redis4}) provide the constitutive equations for an active fluid and thus take no
account of the passive elastic response of a muscle fiber. Such a response, however, is expected in a passive 
fiber, $\Delta\mu=0$, due to the elastic elements of sarcomeres, which help to maintain structural integrity. Consequently,
on large time scales, one
expects an elastic response also in the active case. The short time behavior in the passive case 
is somewhat more subtle. The motor molecules constantly detach from 
actin filaments and rebind. This dynamics leads to a viscous response of the system~\cite{tawa91a,tawa91b}.
Therefore, the passive system shows visco-elastic behavior. Note, that on very short time scales, bound
motors will lead to an additional elastic response. We will neglect this contribution and consider only one relaxation time.
In contrast to theories of active polar gels~\cite{krus04,krus05a}, we thus have viscous behavior on short and elastic
behavior on long time scales. 
We will model this behavior of the passive system by the Kelvin-Voigt model of a linear spring and a dashpot acting 
in parallel, such that the elastic and the viscous stresses simply add. The elastic part is given by $\sigma^\text{el}=E \rho / \rho_0$,
where $\rho_0$ is the equilibrium density and $E$ the elastic modulus. The elastic part must be added to the reactive
component of the stress in Eq.~(\ref{eq:redis3}).

For the analysis of the dynamic equations (\ref{eq:redis1})-(\ref{eq:redis4}), we take a dependence of the active
stress on the density $\rho$ into account: As the overlap between the actin filaments and the motor filaments
increase, the active stress changes for constant $\Delta\mu$. We therefore expand $\zeta$ in terms of $\rho$, while
keeping only terms up to first order,  $\zeta=\zeta_0+\zeta_1 \rho$. In the case that the only external force 
results from friction with the surrounding fluid, we write $\partial_x f_{\text{ext}}=-\eta_e \partial_t \rho / \rho_0$, where $\eta_e$
is the corresponding friction coefficient. Combining this expression with the dynamic equations for $\rho$, the time
evolution of deviations $\rho$ from the equilibrium distribution $\rho_0$ is determined up to linear order by
\begin{equation}
\label{eq:hydroeqnofmotion}
\left( E + \rho_0\:\Delta\mu\:\zeta_1\right) \: \partial_x^2\rho - \xi \: \partial_t\partial_x^2 \rho = \eta_\text{e}\:\partial_t\rho \quad.
\end{equation}
If $E + \rho_0\:\Delta\mu\:\zeta_1 < 0$, a perturbation of the equilibrium state will thus grow leading to a 
contracted state. The condition states essentially that the active stresses must be larger than the passive elastic
stresses. Note, however, that the dynamic equation for the fiber density does not allow for wave solutions. 

\section{Microscopic model of muscle fibers}

As we have seen in the previous section, the oscillation mechanism of muscle fibers involves non-hydrodynamic modes.
In this section, we will propose a microscopic model of the dynamics of muscle fibers. First, we present 
the dynamics of a half-sarcomere and show that it is able to oscillate spontaneously. We 
then study the dynamics of a chain of such elements and compare it to the experimental observations on muscle fibers.
Finally, we obtain the continuum limit of the chain and compare it to the description developed in the previous 
section. 

\subsection{The half-sarcomere}

Muscle fibers are periodic structures. The elementary units are sarcomeres as illustrated in Fig.~\ref{fig:schemasarcomere}.
A sarcomere consists of filaments of myosin-II motors and actin filaments with their plus-ends pointing outwards.
By activation of the motors, the structure contracts. In addition to the active acto-myosin system, there are passive
elastic elements affecting the dynamics of a sarcomere. Elasticity results from structural elements like the Z-disc,
to which actin filaments are attached with their plus-ends, or from titin, a molecule extending through the whole length
of a sarcomere and preventing it from falling apart upon stretching. From a dynamic point of view, the two halves of a 
sarcomere are identical, such that half-sarcomeres can be considered as the elementary units of a muscle fiber. 
We are now going to specify the dynamics of this structure. 

\subsubsection{The dynamic equations.}
\label{sec:dyneq}

\begin{figure}[tbp]
\includegraphics[width=0.5\textwidth]{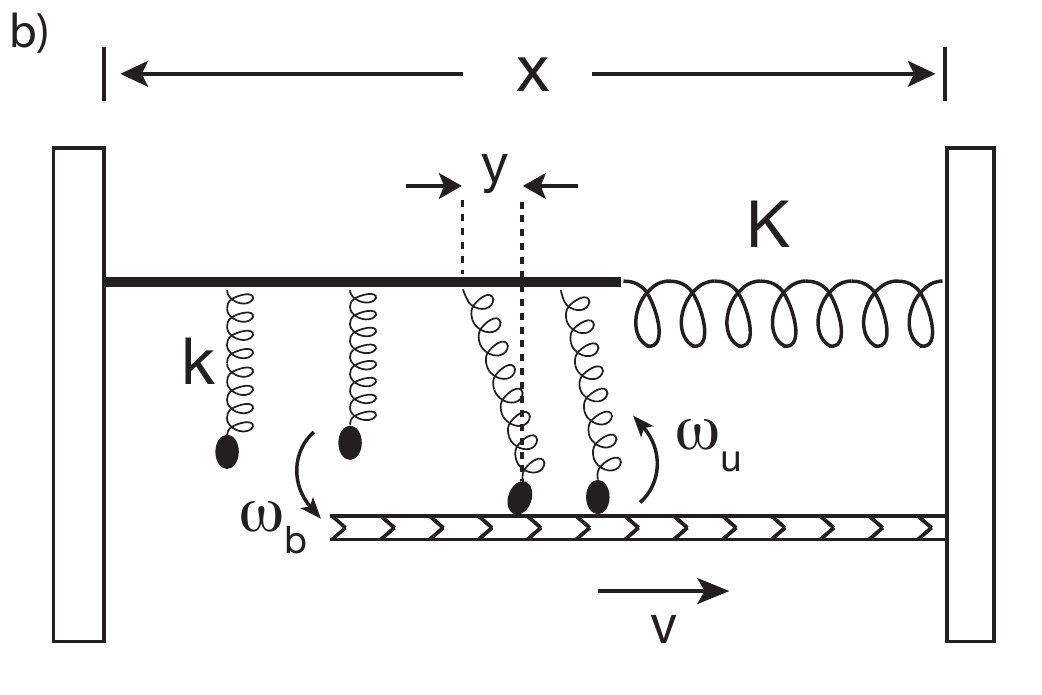} 
\caption{\label{fig:schema}Schematic representation of the half-sarcomere model. Motors are linked to a common
backbone by springs of stiffness $k$ and extension $y$. Motors bound to the polar filament advance at velocity $v$, 
the corresponding binding and unbinding rates are $\omega_b$ and $\omega_u$, respectively. The elasticity provided 
by structural elements of a half-sarcomere is lumped into a spring with stiffness $K$. The extension of the element
is $x$.}
\end{figure}
We approximate a half-sarcomere by the structure illustrated in Fig.~\ref{fig:schema}.
It consists of motors moving along a polar filament and a linear spring of stiffness $K$. The motors are attached 
to a common backbone by springs of stiffness $k$ and extension $y$. The motor filament as well as the polar
filament are effective structures that result from averaging the parallel filaments in a sarcomere in the direction
perpendicular to the sarcomere extension. The whole structure is immersed in a fluid of viscosity $\eta$.

A motor in the overlap region of the motor 
filament and the polar filament stochastically binds to and unbinds from the polar filament with rates 
$\tilde{\omega}_{\text{b}}$ and $\tilde{\omega}_{\text{u}}$, respectively. The binding and unbinding rates 
depend in general on the force applied to the motor. We assume that the force dependence is restricted to the 
unbinding rate. Motivated by Kramers' rate theory we write 
$\tilde{\omega}_{\text{u}}=\omega_{\text{u}}^{0}\,\exp\left\{|f|a/k_{\text{B}}\,T\right\}$, where $a$ is a microscopic 
length scale. The effective binding and unbinding rates used here, can be related to the binding and unbinding rates
of individual motor molecules. If there are $M$ motors in a sarcomere cross-section, then the effective binding and
unbinding rates are, respectively, $\omega_{\text{b}}=M\;\tilde{\omega}_{\text{b}}$ and $\omega_{\text{u}}=
M\;\tilde{\omega}_{\text{b}}\;/\ [ \left(\tilde{\omega}_{\text{b}}/\tilde{\omega}_{\text{u}}+1\right)^{M}-1 ]$~\cite{klum05}.
The averaging and the determination of the parameters for a sarcomere are performed in \ref{app:parameters}.
  
Motors bound to the filament move 
directionally on the polar filament such that the half-sarcomere contracts. For simplicity, we assume a linear 
force-velocity relation $v(f) = v_{0}\,(1-f/f_{0})$. Here, $f=k\,y$ is the force acting on a motor, $f_{0}$ is the stall force 
at which the motor stops walking, and $v_{0}$ is the velocity of an unloaded motor. Implicitly, the linear force-velocity 
relation assumes processive motors, whose mean path bound to a filament noticeably exceeds the step size of a 
single motor. While individual myosin-II molecules are non-processive, the ensemble of motors in a cross-section
can be described by an effective motor that is processive, see \ref{app:parameters}. Unbound motors diffuse in 
the potential provided by the spring which connects them to the backbone. In the following we will refer to effective
motors as motors.

The derivation of the dynamic equations for the half-sarcomere length follows closely the procedure introduced in
Ref.~\cite{gril05}, where mitotic spindle oscillations are studied. The elastic force is given by 
$f_{\text{e}}=K(x-L_{0})$, where $L_0$ is its rest length and $x$ is the length of the element. Let $y_n$ denote the 
extension of the spring linking motor $n$ to the backbone. The total motor force then is 
$f_{\text{m}}=k\sum_{n=1}^{N}\sigma_{n}\,y_{n}$. Here, $N$ is the number of motors  and $\sigma_{n}=1$ if motor 
$n$ is bound and $0$ otherwise. The extension $y_n$ changes according to
\begin{equation}
\label{eq:ydot}
\dot y_n=v_n+\dot x\quad,
\end{equation}
where $v_n$ is the velocity of motor $n$ on the filament and $\dot x$ describes changes of the length of the 
half-sarcomere element. 

In the following, we will use a mean-field approximation, which consists in assuming that all motors in the half-sarcomere
have the same spring extension, $y_n=y$ for all $n$. The mean position is determined through $y=\dot y/\omega_{\text{u}}$. 
The fraction of bound motors is denoted by $Q$. Assuming fast relaxation of unbound motors, their distribution equals the 
equilibrium distribution and the time evolution of $Q$ can be shown to obey~\cite{gril05}
\begin{equation}
\label{eq:qdot}
\dot Q=\omega_{\text{b}}-(\omega_{\text{b}}+\omega_{\text{u}})\,Q\quad,
\end{equation}
while the total motor force is given by $f_{\text{m}} = N(x)\,Q\,k y$.
Here, $N(x)=(\ell_{\text{f}}+\ell_{\text{m}}-x)/\Delta$ is the number of motors in the overlap region of
the filament and the motor backbone of lengths $\ell_{\text{f}}$ and $\ell_{\text{m}}$, respectively,
while, $\Delta$ is the distance between adjacent motors. 

The elastic forces as well as the active forces generated by motors are balanced by friction and possibly by an external
force:
\begin{equation}
\label{eq:hsforcebalance}
f_{\text{e}}+f_{\text{m}}=f_{\text{f}}+f_{\text{ext}}\quad.
\end{equation}
The Reynolds numbers associated with flows generated by shortening of the half-sarcomere is low, such that inertial
effects can be neglected. Accordingly, for the friction force $f_{\text{f}}$ we write $f_{\text{f}}=-\xi \dot{x}$, where $\xi$
is an effective friction coefficient which depends on the viscosity $\eta$ of the surrounding medium. Note, however,
that there is also a contribution of the motors to friction~\cite{tawa91a,tawa91b}, see \ref{app:parameters}. 
This completes the specification of the dynamics of the model half-sarcomere.

\subsubsection{The phase diagram.}

The dynamic equations allow for a stationary state $(x_{0},Q_{0},y_{0})$, where the active force of the motors is 
balanced by the elastic forces. Note, that for too strong motors, that is, if $v_0$ or $\tilde\omega_b$ are too large, this 
state is unphysical as 
the structure shortens to values below the length of the actin or the motor filament, 
$x_0<\max\{\ell_{\text{m}},\ell_\text{f}\}$. A linear
stability analysis of the stationary state yields ${y}_{0}\,\omega_{\text{u}}'({y}_{0})>\omega_{\text{b}}+
\omega_{\text{u}}({y}_{0})$ as a necessary condition for an instability. This implies that the force dependence
of the motor binding kinetics is essential for oscillations. A second necessary condition for an oscillatory 
instability is $Q_{0}\,k\,{y}_{0}<K\,\Delta$. This means that the force generated in the stationary state by an 
additional motor in the overlap region of the polar and the motor filament is smaller than the corresponding
change in the elastic force.

\begin{figure}[tbp]
\includegraphics[width=0.5\textwidth]{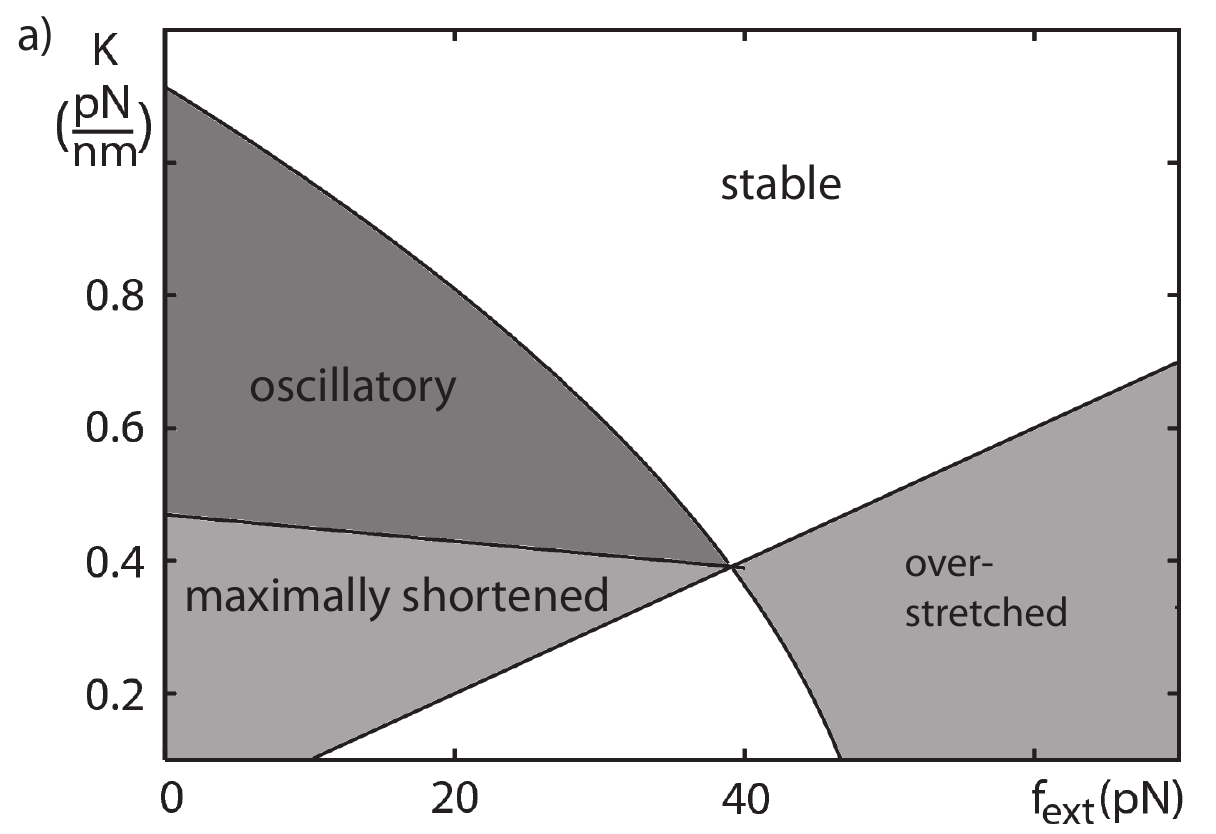} 
\includegraphics[width=0.5\textwidth]{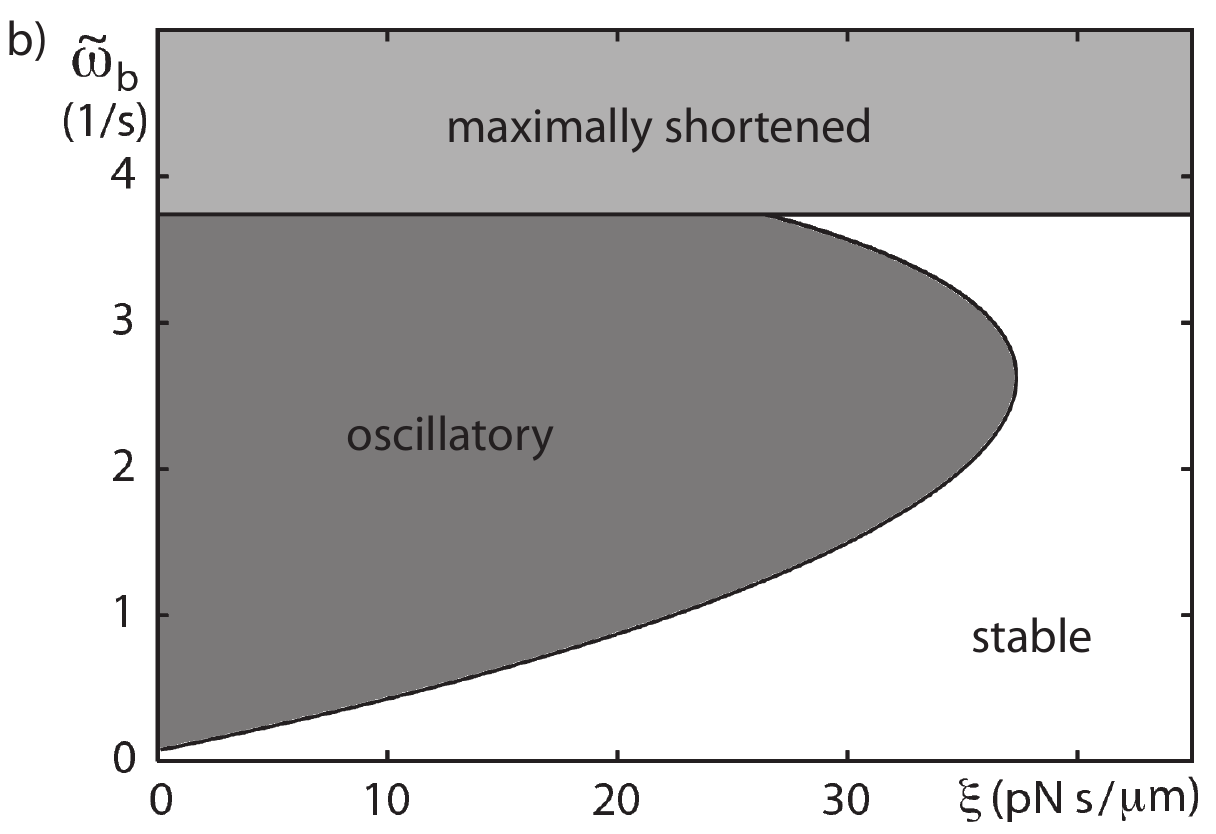} 
\caption{\label{fig:linstab}Phase diagram of  a half-sarcomere. a) Dynamic states as a function of 
the external force $f_\text{ext}$ and the stiffness $K$. Regions of a stable stationary state are white. In the dark gray 
region, the system oscillates spontaneously. In the light gray region, the stationary state is unphysical, implying that the
element either fully contracts or over-stretches. b) Dynamic states as a function of the effective friction coefficient
and the binding rate of individual motors. In (a) and (b) all other parameters are as in \ref{app:parameters}.}
\end{figure}
Figure~\ref{fig:linstab} presents the regions of stability of the stationary state for different cuts through the parameter
space. In Fig.~\ref{fig:linstab}a, the strength of the external force and of the elastic element $K$ are varied.
For sufficiently small external forces and stiffness $K$, the motors maximally shorten the element. As the
stiffness is increased, the system starts to oscillate spontaneously. Upon a further increase, an inverse Hopf-bifurcation
occurs and the stationary state is stable. Eventually, in the limit of large $K$, the system will be forced to assume
a length close to the rest length $L_0$. If instead the external force $f_\textrm{ext}$ is increased, a region
of stable physical stationary states is reached. Beyond a critical value of the external force, though, the system 
is over-stretched and behaves as a simple elastic element with stiffness $K$ as the motors cannot interact with the
polar filament. Note, that since the elastic elements of a real sarcomere are non-linear, a change in $f_\text{ext}$
will affect also the value of $K$ that we use in the model.

In Figure~\ref{fig:linstab}b,
we present the phase diagram resulting from changes in the motor activity and in the viscosity of the surrounding 
fluid. The latter influences the value of the effective friction constant $\xi$. Changes in the motor activity affect
several parameters. An important parameter is the binding rate $\tilde{\omega}_{\text{b}}$ of individual motors.  
The diagram shows that a decrease of the effective friction constant can induce 
oscillatory behavior. For sufficiently low values of $\xi$, an increase of $\tilde{\omega}_{\text{b}}$ can first lead to an
instability of the stationary state, while for too large values it is stable. This is reminiscent of the behavior predicted 
for spindle oscillations by a similar model~\cite{gril05}, and for which there is experimental evidence~\cite{pecr06}.
Note, that
due to the mechanism studied in Ref.~\cite{tawa91a,tawa91b}, changes in $\tilde{\omega}_{\text{b}}$
also affect the value of $\xi$.

\begin{figure}[tbp]
\includegraphics[width=0.3\textwidth]{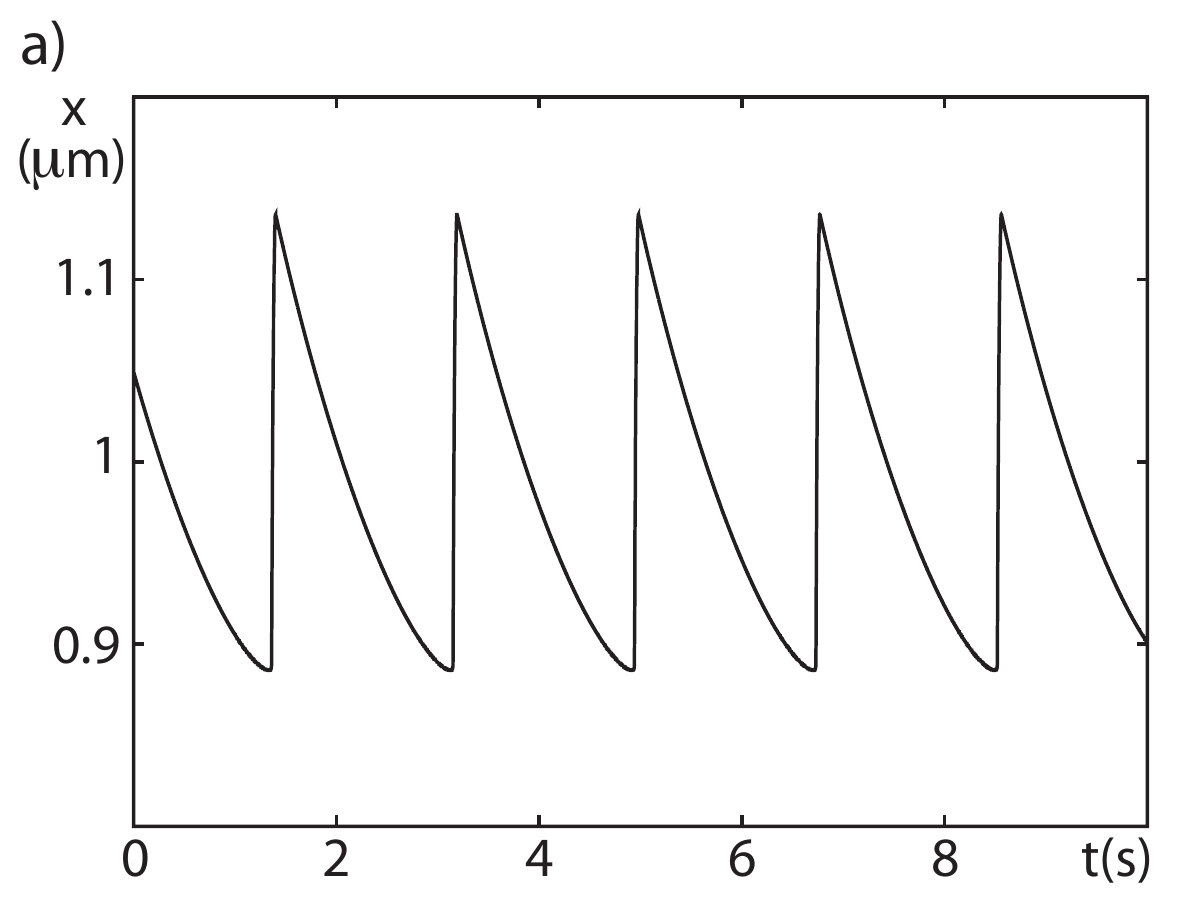} 
\includegraphics[width=0.3\textwidth]{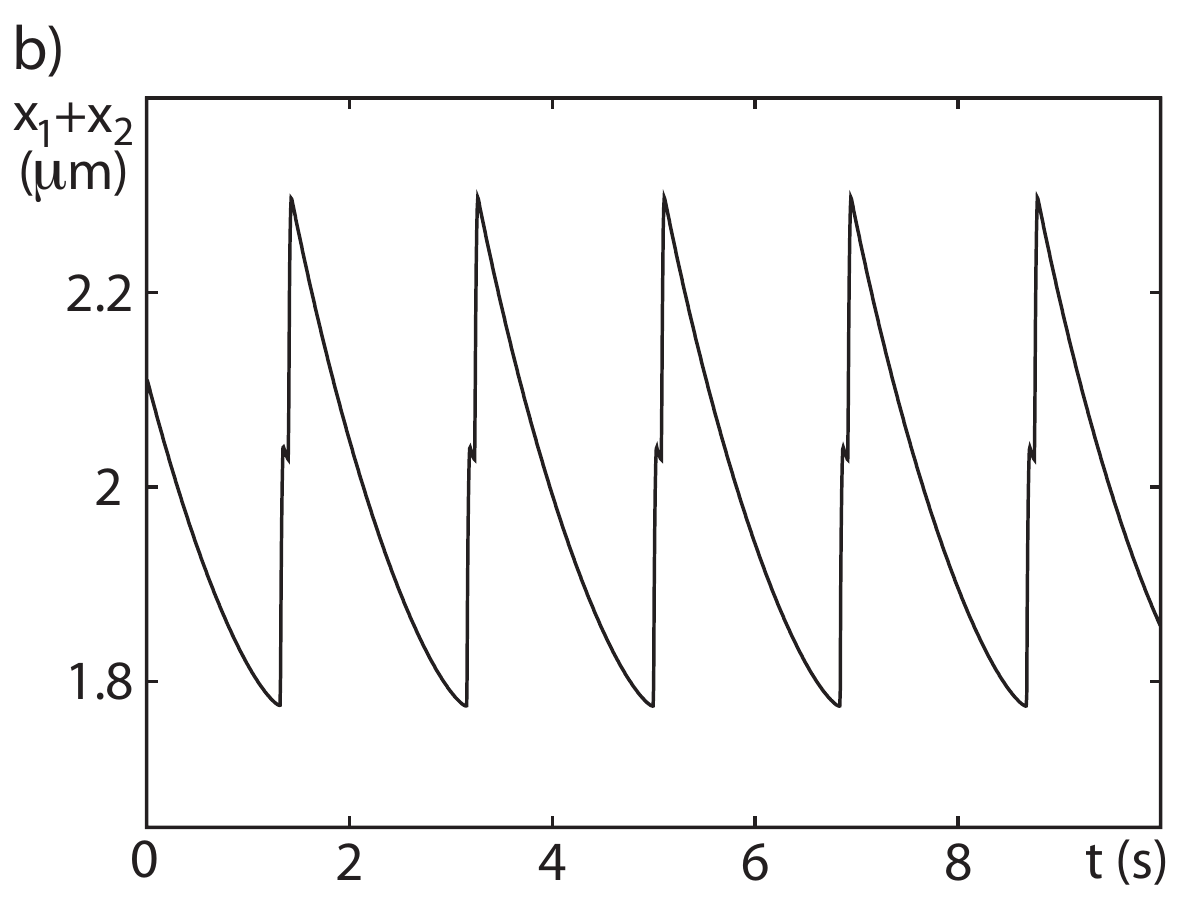} 
\includegraphics[width=0.33\textwidth]{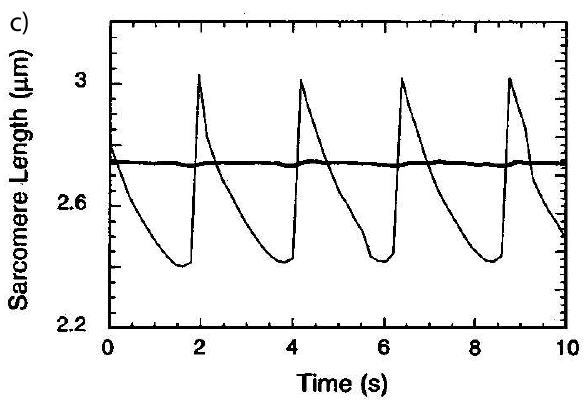} 
\caption{\label{fig:hsosc}Sarcomere oscillations. a) Oscillations in the extension $x$ of a half-sarcomere as a function
of time generated by 
Eqs.~(\ref{eq:ydot})-(\ref{eq:hsforcebalance}). b) Oscillations in the extension $x_1+x_2$ of two coupled half-sarcomeres
as a function of time. In (a) and (b)
the parameters of \ref{app:parameters} have been used.
c) Averaged oscillations of one sarcomere of skeletal muscle. Adapted from \cite{yasu96}. The numerical
solutions shown in (b) and the experimental oscillations in (c) have the same temporal period and similar 
amplitudes. In all cases the functional form indicates a phase of relative slow contraction followed by a 
fast extension.}
\end{figure}
An example of an oscillation of a half-sarcomere is shown in Fig.~\ref{fig:hsosc}a. It shows the characteristic saw-tooth
shape which is observed experimentally: a slow contraction of the element is followed by a rapid expansion.
As the motors shorten the element, the elastic force increases and therefore the force on the motors. This in turn 
increases the unbinding rate of motors. As a few motors detach from the filament, the remaining motors experience an 
even higher force, such that an avalanche of motor unbinding events occurs. The elastic element stretches the linker 
again and the motors rebind, the cycle can repeat.

\subsection{A chain of half-sarcomeres}
\label{sec:chain}

We now consider a chain of half-sarcomeres as studied in the previous section. It will be shown that the
strong coupling between half-sarcomeres in a chain leads to traveling wave solutions that share essential 
features with waves observed experimentally in muscle fibers.

Consider a chain of $S$ half-sarcomeres, where the right end of a half-sarcomere is simultaneously the left
end of the subsequent element. The right end of half-sarcomere $i$ in the chain is at position $z_i$,
the left end of the first-sarcomere is fixed at $z_0=0$. Force balance of the half-sarcomere ends yields:
\begin{eqnarray}
\label{eq:scforcebalance1}
f_{\text{f,1}} &=& \left(f_{\text{e,1}}+f_{\text{m,1}} \right) - \left(f_{\text{e,2}}+f_{\text{m,2}} \right) \\
\label{eq:scforcebalance2}
f_{\text{f,j}} &=& 2 \left(f_{\text{e,j}}+f_{\text{m,j}} \right) - \left(f_{\text{e,j-1}}+f_{\text{m,j-1}} \right) - \left(f_{\text{e,j+1}}+f_{\text{m,j+1}} \right)\\
\label{eq:scforcebalance3}
f_{\text{f,S}} &=& 2 \left(f_{\text{e,S}}+f_{\text{m,S}} \right) - \left(f_{\text{e,S-1}}+f_{\text{m,S-1}} \right) + f_\text{ext} \quad,
\end{eqnarray}
with $j=2 \dots S-1$. Here, $f_\text{f,j}=-\xi{\dot x}_j$, where $x_j=z_j-z_{j-1}$ and $j=1,\ldots,S$, denotes the friction
force in the $j$th half-sarcomere. The forces $f_\text{e,j}$ and $f_\text{m,j}$ are, respectively, the elastic 
and the motor force of half-sarcomere $j$. The expressions for $f_\text{e,j}$ and $f_\text{m,j}$ are 
identical to those of an individual half-sarcomere. The dynamic equations for the extension $y_j$ and the fraction $Q_j$
of bound motors in half-sarcomere $j$ are obtained from Eqs.~(\ref{eq:ydot}) and (\ref{eq:qdot}), respectively, where $x$,
$y$, and $Q$ are replaced by $x_j$, $y_j$ and $Q_j$. 

For the parameters given in \ref{app:parameters}, a chain of $S=2$ half-sarcomeres, i.e.,  sarcomere, the asymptotic
dynamic state is presented in Fig.~\ref{fig:hsosc}b and compares nicely to experimental observations~\cite{yasu96},
see Fig.~\ref{fig:hsosc}c. While a half-sarcomer essentially only shows two states, a stationary and an oscillatory, the dynamics of a sarcomere is richer. Oscillations of different symmetry classes can be identified as is reported in
Ref.~\cite{guntUP} where hydrodynamic interactions between half-sarcomeres have been included. A detailed study 
of the bifurcation diagram of a sarcomere will be presented elsewhere.

\begin{figure}[tbp]
\includegraphics[width=0.5\textwidth]{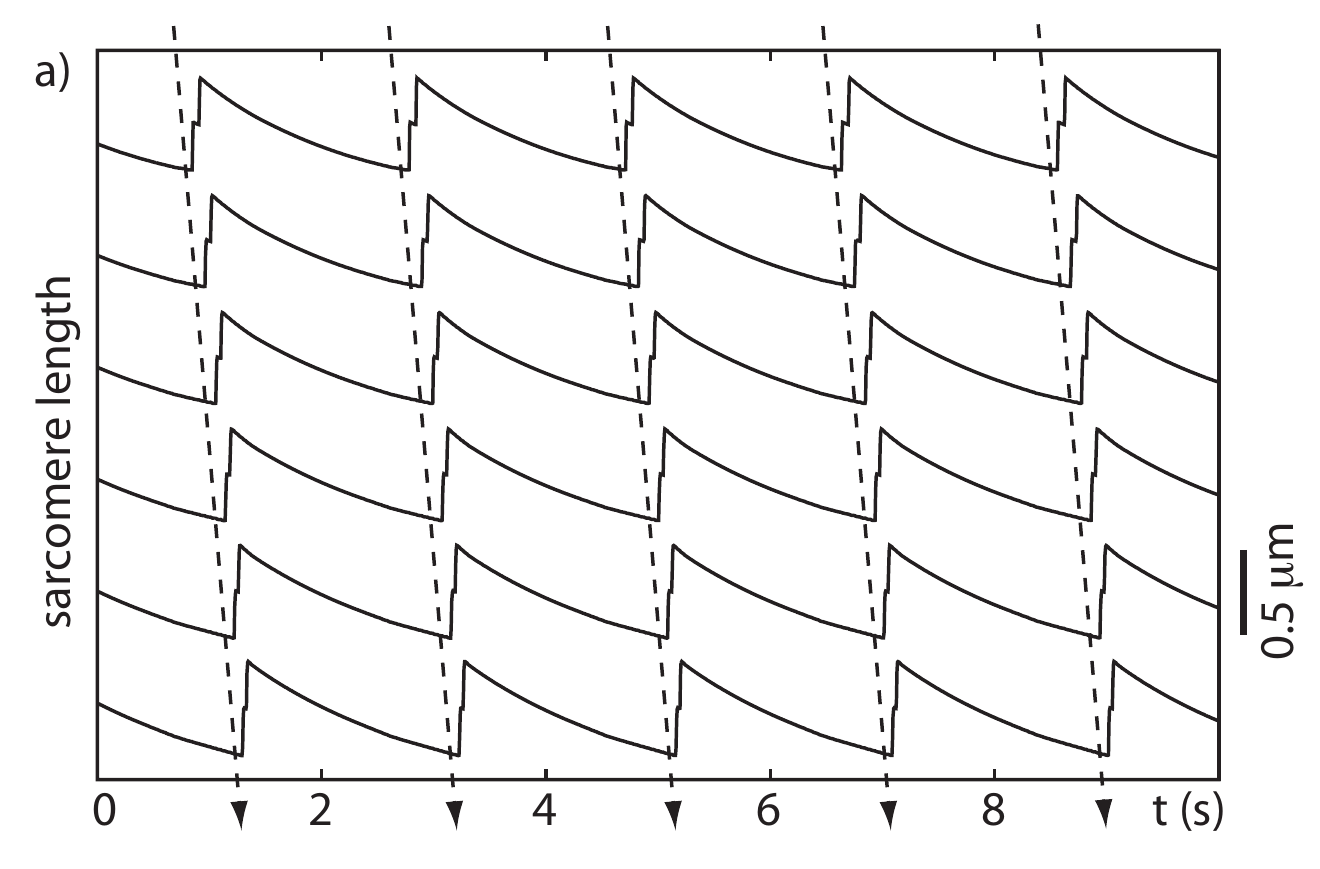} 
\includegraphics[width=0.5\textwidth]{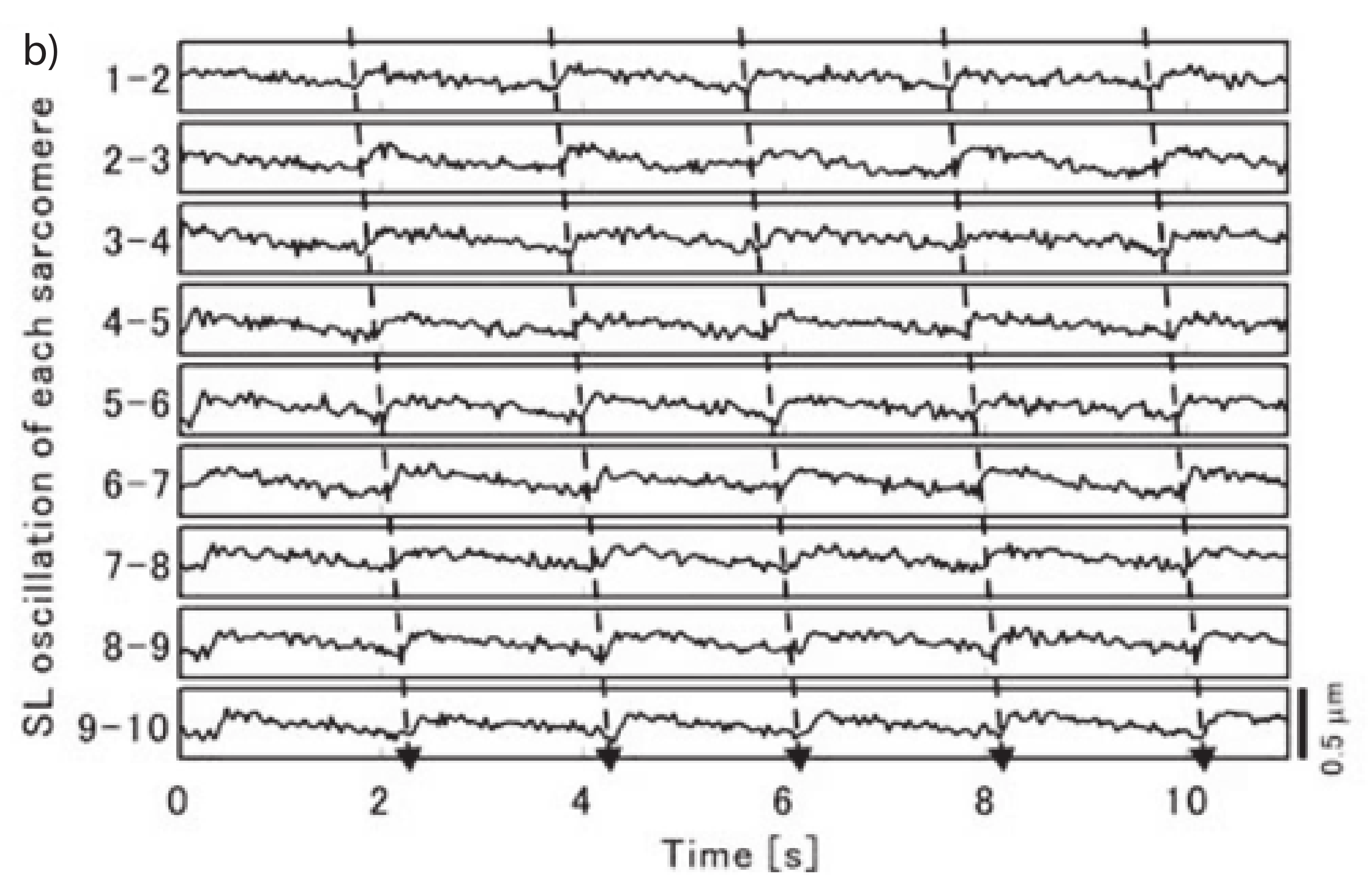} 
\caption{\label{fig:waves}Spontaneous travelling waves. a) Wave along a chain of $S=20$ half-sarcomeres as described 
in Sec.~\ref{sec:chain}. The extensions of the six central sarcomeres are depicted successively on top of each other, with 
the left-most being on top. The dashed arrows connect the positions of the maximally contracted state in adjacent 
sarcomeres. Parameters are as in \ref{app:parameters}. b) Wave observed in a cardiac muscle fiber. Modified 
from Ref.~\cite{sasa05}.}
\end{figure}
In Figure~\ref{fig:waves} a spontaneous wave along a chain of 20 half-sarcomeres is presented for the parameter
values given in \ref{app:parameters} and compared to an experimentally observed wave in a cardiac muscle
fiber~\cite{sasa05}. In both cases a relaxation wave travels from the left to the right. The temporal period of individual 
sarcomeres is the similar in the simulation and the experiment, the corresponding oscillation amplitude somewhat
larger in the theory. The relative phase shift between the oscillations in adjacent sarcomeres differs by less than 10\%
between theory and experiment. 

Note, that experimentally, cases are also observed, where the relaxation wave starts at the right of the fiber rather
than at the left, leading to a wave traveling from right to left. In still other cases, the wave is initiated at some point 
along the fiber, traveling to the right and to the left starting form that point. The same is observed in numerical solutions 
of a chain of half-sarcomeres: depending on the initial conditions the wave can start anywhere along the chain.  
In addition to regular waves, for other parameter values we also find states where individual sarcomeres oscillate,
but no coherent relaxation wave is formed.

\subsection{The continuum limit}

In order to get further insight into the spontaneous waves and to compare the model presented in this section
with the phenomenological approach presented in Sect.~\ref{sec:hydrodynamic}, we now study the continuum 
limit of an infinite chain. It will be shown, that in this limit and neglecting the binding dynamics of the motors,
the hydrodynamic description is obtained.

In the continuum limit, we will neglect all non-linear terms. Thus we first linearize Eqs.~(\ref{eq:ydot}), (\ref{eq:qdot}) 
and (\ref{eq:scforcebalance2}) with respect to the stationary state ($x_{0},Q_{0},y_{0})$. In the linearized 
Eq.~(\ref{eq:scforcebalance2}), we then approximate the differences $z_j-z_{j-1}\simeq x_0\partial_z x + x_0$, where
$z$ is now the coordinate along the chain.  
The gradients in $x$ are just the strain in the chain and can be linked to the density $\rho(z)$ along the chain. Consider
a piece of length $L$ of the chain. On one hand, changing the density from its stationary value $\rho_0$ to $\rho$ 
changes the mass in this piece as  $\int_L(\rho_0-\rho) dz$. On the other hand, the change in density can be 
linked to a change in the distance between the particles (the end-points of the half-sarcomeres) according to
$\rho_0[x(z_L)-x(z_0)]=\rho_0\int_L \partial_z x\,dz$, where $z_0$ and $z_L$ are the end points of the piece.
Since the piece was arbitrarily chosen, we find $\partial_z x=1-\rho/\rho_0$.

Eliminating the equation for $y$, we then arrive at
\begin{eqnarray}
\label{eq:drhodtcont}
\xi\partial_t\rho & = & \left[c_1+c_2\partial_t\right]\partial_z^2\rho - c_3\partial_z^2Q\\
\label{eq:dQdtcont}
\partial_t Q & = & c_4\partial_t\rho - c_5 Q\quad.
\end{eqnarray}
Here, the constants are related to the microscopic parameters via
\begin{eqnarray}
\label{eq:c1}
c_1 & = & [K-kQ_0y_0/\Delta]x_0^2\\
\label{eq:c2}
c_2 & = & kN(x_0)Q_0x_0^2/[\omega_u^\prime(y_0)y_0+\omega_u(y_0)+kv_0/f_0]\\
c_3 & = & \rho_0x_0N(x_0)ky_0\\
c_4 & = & Q_0\omega_u^\prime(y_0)x_0/[\rho_0(\omega_u^\prime(y_0)y_0+\omega_u(y_0)+kv_0/f_0)]\\
c_5 & = & \omega_b+\omega_u(y_0)\quad.
\end{eqnarray}
In the case of a stationary fraction $Q$ of bound motors, Eq.~(\ref{eq:drhodtcont}) has the same form as the
hydrodynamic equation of motion (\ref{eq:hydroeqnofmotion}) and relations (\ref{eq:c1}) and (\ref{eq:c2})
relate the phenomenological parameters to microscopic parameters. We again find, that the dynamics of the
fraction $Q$ of bound motors, which clearly is not a hydrodynamic mode, is essential for obtaining waves 
along the a chain of half-sarcomeres. A linear stability of the homogenous stationary state of  Eqs.~(\ref{eq:drhodtcont})
and (\ref{eq:dQdtcont}) yields the same necessary condition as above for oscillatory solutions, 
$Q_{0}\,k\,{y}_{0}<K\,\Delta$ as found above.

\section{Discussion}
In the preceding sections, we have presented theoretical descriptions of the dynamics of muscle fibers.
Using a hydrodynamic approach where the system is described as a one-dimensional complex fluid close
to thermodynamic equilibrium, we obtained a bound on the system's activity to obtain contraction. Based 
upon a microscopic model of half-sarcomeres, we found oscillatory states that correspond to traveling waves
along the fiber. The essential ingredient underlying this dynamic behavior is contained in a force dependence
of the binding-unbinding kinetics of myosin motors to actin filaments. Spontaneous oscillations of sarcomeres
and contraction waves along muscle fibers are observed experimentally and have been studied 
intensively~\cite{okam88,anaz92,ishi93a,ishi93b,yasu96,fuku96,fuji98,sasa05}. In the following paragraphs,
we will compare experimental findings to our theoretical results.

First of all, sarcomeres have been observed to spontaneously oscillate under constant non-physiological 
conditions~\cite{okam88}. They consist of a slow contraction and a fast expansion phase, leading to a
saw-tooth pattern of the sarcomere extension vs. time curve. It was shown experimentally that the oscillations
are not induced by resonances with an external load, leading to the conjecture that spontaneous oscillations
should be possible also in the absence of an external load~\cite{fuji98}. Our (half-)sarcomere model generates
behavior that is certainly consistent with these results and conclusions. 

By choosing parameter values that
are compatible with the values known from muscle and muscle myosin, the oscillatory solutions we find 
match the experimentally observed semi-quantitatively. Parameters could always be fitted such that the
experimentally observed period and amplitude of the oscillations match exactly. Note, however, that these
differ for different kinds of muscle. We therefore chose one typical set of parameters as explained in \ref{app:parameters}
without aiming at quantitatively matching a particular experiment. 

Parameter values have been systematically varied in experiments. Unfortunately, variations in the chemical
composition of the buffer solution, are not readily translated into changes of the parameter values we employ
in our model. This holds in particular for the activity of the motors. In the buffers studied experimentally, motors
are always only partially activated. The degree of the activation, however, is not assessed directly. Still, a few
comparisons can be made. First of all, the oscillation frequency does not depend on the applied external
force~\cite{yasu96}. The same holds for the critical frequency at the onset of the oscillatory instability in 
the model.  Furthermore, it has been shown that the oscillatory regime is an intermediate state between relaxation 
and contraction~\cite{okam88,ishi93a} that can be continuously connected by changing the degree of motor 
activity~\cite{fuku96}. Changes in the degree of motor activity, can in our model for example be described by
increasing the binding rate $\tilde\omega_b$ of motors to the polar filament.  In that case, we also observe a transition
from a relaxed state, to an oscillatory state and then a maximally contracted state, see Fig.~\ref{fig:linstab}b.

Experimentally, spontaneous oscillations are observed only if a moderate external force is applied~\cite{anaz92}.
At first sight, this might be incompatible with the phase diagram presented in Fig.~\ref{fig:linstab}a. Indeed, if only
the external force is varied, then the system can at best be driven form an oscillating state to a non-oscillating one.
However, one has to keep in mind, that we have made the assumption of linear elastic elements. In a sarcomere,
though, the elasticity is non-linear. While for the oscillations it is not essential to keep the non-linearity, it implies
that with the application of an external forces also the value of the effective stiffness $K$ changes (it might either
increase or decrease). By describing a curvilinear path in the ($f_\text{ext}$,$K$)-plane when changing $f_\text{ext}$,
the experimental observation and the theoretical results can be compatible. Furthermore, changes in the external
force might affect other model parameters, for example, through the mechanism of stretch activation.
 
As already mentioned in the previous section, all the kinds of traveling waves observed in experiments have their
counter-part in solutions to the chain equations. In particular, the propagation speed in terms of relative phase shifts
between adjacent sarcomeres compares nicely.

In our analysis, we have neglected the effects of noise in our system. It has been shown in Ref.~\cite{gril05}, where
a similar approach was followed to describe oscillations of mitotic spindles during asymmetric cell division, that 
the mean-field analysis faithfully reproduced the features of the stochastic system. The same can be expected to
hold in the present case. 

In summary, we have presented a first semi-quantitative analysis of spontaneous oscillations of muscle fibers.
It will now be interesting to design experiments that could specifically change parameters of the model and
thereby test the importance of a force-dependent binding-unbinding kinetics of molecular motors for the
observed oscillations. Due to the wide-spread appearance of motor oscillations, a quantitative
combined experimental and theoretical study will likely help us to understand various cellular processes.

\appendix

\section[]{Parameters}
\label{app:parameters}
In this appendix, we discuss the parameter values that we have used in our numerical solutions of the
dynamic equations presented in Sect.~\ref{sec:dyneq}. As the parameter values differ for the various
kinds of muscle used in experiments, we will present here typical values rather than trying to argue why
a specific value has to be chosen for describing a particular experiment.

Let us start by geometric considerations. A typical rest length of an inactive sarcomere is $2.5\mu$m~\cite{albe02}, 
so we take $2L_0=2.5\mu$m, where $L_0$ is the rest length of the elastic element of a half-sarcomere. Typical
lengths of actin filaments are $\ell_{\text{f}}=0.6\;\mu$m and $\ell_\text{m}=1.5\;\mu$m for myosin filaments, the
latter consisting of about 300 Myosin-II molecules~\cite{albe02}. Correspondingly, the average distance between 
two adjacent motors on a myosin filament is $\Delta=5$nm. 

In a muscle fiber, the approximately 1000 myosin filaments are arranged in a hexagonal lattice interdigitating 
with a hexagonal lattice of actin filaments. In a slice of thickness $\Delta$ perpendicular to the orientation of the
filaments, however, only a fraction of the motors can actually bind to an actin filament. This is because myosin
binding sites are evenly spread on an actin filament every 37nm. Assuming that a motor head can bind within 4nm 
around its equilibrium position with respect to the motor filament, we arrive at about 100 motors that actually
can bind in a slice of thickness $\Delta$.

As mentioned in Sect.~\ref{sec:dyneq}, an ensemble of $M=100$ non-processive Myosin-II can  be considered
as one effectively processive motor. The time an individual motor spends bound to an actin filament is about 
5ms~\cite{howa01}. As in the experiments motors are only partially activated~\cite{okam88,fuku96}, we choose
here $\tilde{\tau}_{\text{b}}=1/\omega_u^0=30$ms. The binding rate can then be inferred from the duty ratio $r$, which 
gives the fraction of time a motor is attached to a filament during a whole cycle. We choose $r=0.09$ which is in 
the middle of the predicted range~\cite{howa01}. From the values of individual motors, the parameters characterizing
the effective motor can be inferred~\cite{klum05}. For the effective motor, we furthermore choose a stall force $f_0=4$pN
and a load-free velocity $v_0=0.4\mu$m/s. These values are lower than what might be obtained from single molecule
experiments and again account for the only partial activation of motors in the experiments.

The elastic components entering our model are experimentally hard to assess. Single molecule experiments with 
Myosin-II suggest a stiffness of the linker of the head to the motor filament of the order of 
$k=4$pN/nm~\cite{howa01,schw02}. We are not aware of measurements of $K$. Contributions come from titin 
molecules~\cite{tskh03} the Z-discs and other elements. We choose $K=0.5$pN/nm. Similarly, the microscopic
length scale $a$ is unknown. It must be of molecular dimensions and we choose $a=3\;$nm. Experiments are
carried out at room temperature, such that $k_{\text{B}}T=4$pN nm.

Finally, friction in muscle fibers results from protein-protein friction as well as from hydrodynamic friction. Their effects
are lumped into the parameter $\xi$. Assuming a viscosity of ten times the viscosity of water $\eta_\text{water}\approx 
10^{-3}\;$Pa s, we use for the effective friction coefficient $\xi=10\;$pN s$/\mu$m.

\end{document}